\documentclass[aps,showpacs,preprint,superscriptaddress]{revtex4-1}
\usepackage{graphicx}
\usepackage{subfigure}
\usepackage{float}
\usepackage{amsmath}
\usepackage{amsfonts}
\usepackage{color}
\usepackage{bm}
\usepackage{bbm}
\usepackage{txfonts}
\usepackage{array}
\usepackage{amssymb}

\begin{document}
\title{Dynamically assisted pair production enhancement by combined multiple potentials}
\author{Lie-Juan Li}
\affiliation{School of Mathematics and Physics, Lanzhou Jiaotong University, Lanzhou 730070, China}
\affiliation{Key Laboratory of Beam Technology of Ministry of Education,
School of Physics and Astronomy, Beijing Normal University, Beijing 100875, China}
	
\author{Li Wang}
\affiliation{Key Laboratory of Beam Technology of Ministry of Education,
School of Physics and Astronomy, Beijing Normal University, Beijing 100875, China}

\author{Melike Mohamedsedik}
\affiliation{College of Xinjiang Uyghur Medicine, Hotan 848000, China}

\author{Li-Na Hu}
\affiliation{Key Laboratory of Beam Technology of Ministry of Education,
School of Physics and Astronomy, Beijing Normal University, Beijing 100875, China}

\author{Bai-Song Xie}
\email[Corresponding author. Email: ]{bsxie@bnu.edu.cn}
\affiliation{Key Laboratory of Beam Technology of Ministry of Education,
School of Physics and Astronomy, Beijing Normal University, Beijing 100875, China}
\affiliation{Institute of Radiation Technology, Beijing Academy of Science and Technology, Beijing 100875, China}

\begin{abstract}
We propose a new Sauter-like field model with combinatorial multiple potentials consisting of a deep slow-varying and some shallow fast-varying potentials. The dynamically assisted Sauter-Schwinger effect on the pair production is found by using the computational quantum field theory. The enhanced pair production is found to be significant at about one order increasing for multiple potentials rather than single potential. In case of dominated by Schwinger mechanism, the obvious time effect leads to electrons concentrating at the two edges of the potential, meanwhile, the momentum locates at the zero nearby. In contrary, however, for the multiphoton processes, the pair generation makes the electrons distributing outside the potential and the momentum appearing multiple peaks far away from zero and evenly evolving toward a step-like structure. An interesting finding is that the particles of pair produced in the alternating potential has a quasi-monoenergetic structure compared to the oscillating potential well or/and potential barrier, which is helpful to achieve the high quality positron source.
\end{abstract}

\maketitle

\section{Introduction}
Since that positron was predicted theoretically by Dirac~\cite{Dirac1928} and confirmed experimentally by Anderson ~\cite{Anderson1933}, many classic works had been performed for the problem involving of electronpositron pair creation from the quantum vacuum state triggered by strong fields~\cite{Sauter1931,Heisenberg1936,Schwinger1951}.
Now very useful research methods were developed to investigate this interesting topic, which include the proper time technique~\cite{Tsai1973}, the Wentzel-Kramers-Brillouin (WKB) approximation~\cite{Kim2002}, the worldline instanton technique~\cite{Gies2005,Dunne2005,Dunne2006,Schneider2016}, quantum Vlasov equation solution method~\cite{Alkofer2001,Kluger1991,Abdukerim2017}, the Furry-picture quantization~\cite{Moortgat2009,Aleksandrov2020}, the Dirac-Heisenberg-Wigner formalism~\cite{Hebenstreit2011,Xie2017,Kohlfurst2018,Kohlfurst2020,Liliejuan2021,Mohamedsedik2021}, the computational quantum field theory~\cite{Lv2013a,Wangli2019,Cheng2009,Jiang2013,Krekora2005} and so on. Some reviews and recently works can be seen in publication~\cite{Fedotov:2023ely,Gonoskov:2022hwf}.

In the constant electric field $E$, the pair creation rate is proportional to
$\exp(-\pi E_{\rm{cr}}/E)$, in which the tunneling rate is exponentially suppressed unless the field strength approaches the critical value $E_{\rm{cr}}\approx 1.3\times 10^{16}~\rm{V/cm}$, which corresponds to the laser intensity of $2.3\times 10^{29}~\rm{W/cm^2}$. This electron-positron pair creation process through quantum tunneling is called the Schwinger mechanism~\cite{Schwinger1951}.
Although this prediction has not been directly verified experimentally, the rapid developments of the technology for generating high-power laser pulses continues to encourage theoretical and experimental research, for example, the present maximum intensity achieved is about $10^{23}~\rm{W/cm^2}$~\cite{Yoon2021}.

By using a superintense laser wave propagating through a nuclear Coulomb field, electron-positron pairs can be created through the Bethe-Heitler process~\cite{Muller2003,Augustin2014,Krajewska2021}.
The other routine of electron-positron pair production in strong-field is the Breit-Wheeler process with a high-energy $\gamma$ photon colliding with an ultrastrong laser field~\cite{Krajewska2012,Piazza2016,Titov2018,He2021,Golub2021,Wan2020,Tang2021}.
To our knowledge, the first and also solely experimental positron production observation was performed in Stanford Linear Accelerator Center (SLAC)~\cite{Burke1997} by a collisions of an electron beam of $46.6~\rm{GeV}$ with the terawatt laser pulse at about $10^{18}~\rm{W/cm^2}$ intensity, in which the nonlinear Compton scattering occurs first and then the Breit-Wheeler process creates the positrons.
In addition to nonlinear Breit-Wheeler pair creation, other quantum electrodynamic (QED) processes in the regime of strong fields, such as nonlinear Compton scattering~\cite{Ilderton2020,King2021} and QED vacuum pair generation, also attracted much attention.

For pair production from vacuum under strong background field, various electric fields were
employed to enhance the pair-creation rate. Beside the Schwinger mechanism, some of other
mechanisms were proposed. For an alternating field, although electron-positron pairs can be created by multiphoton processes, the generation rate is exponentially suppressed at low frequencies~\cite{Brezin1970}. Therefore, the increasing observation of pair creation requires either an increase in the field strength, or in the field frequency, or evenly in both of them~\cite{Brezin1970}. In particular, in the subcritical field regime, two mechanisms, Schwinger or/and multiphoton, are strongly weaken. In order to overcome this shortage, the dynamically assisted Schwinger machanism was proposed\cite{Schutzhold2008}, in which it is realized by combing a strong and slow-varying electric field with a weak but fast-varying one. It would lead the tunneling barrier reducing and a drastic enhancement of the electron-positron pair creation rate.

The Sauter potential is a widely used external field model~\cite{Orthaber2011,Wangli2019,Tang2013,Wang2018,Gong2018,Su2020}.
As the extension of the Sauter field, Orthaber $et~al.$ proposed the combination of two Sauter-like laser pulses, both spatially homogeneous~\cite{Orthaber2011}.
It was found that the momentum spectrum has a strong nonlinear behavior under the dynamically assisted Schwinger machanism. The different momentum signatures were also observed.
The dependence of the dynamically assisted Sauter-Schwinger effect
on the temporal shape of the weaker field was investigated by employing a perturbative expansion in terms of the weaker field while treating the strong field non-perturbatively~\cite{Torgrimsson2017}. Since the achievable spatial focusing scale is orders of magnitude larger than the Compton wavelength, many workers ignored the spatial dependence.
However, numerous investigations demonstrated that the electron-positron spectrum is extremely sensitive not only to the temporal but also to the spatial profile of the electric field~\cite{Hebenstreit2014,Liliejuan2021,Mohamedsedik2021}.
Hence, as a first step to approach the realistic field configuration which has more complicated space-time form, the spatial effect of field should be taken into account.

Upon the enhancement of electron-positron pair creation rate under combined Sauter potentials,
recently some effect have been revealed significantly~\cite{Tang2013,Gong2018,Jiang2012}. For example, the frequency modulation effect plays an important role in the enhancement~\cite{Liliejuan2021,Gong2020}. From the analysis of frequency spectrum, strong high-frequency components can improve the pair generation rate, but it is difficult to achieve high intensity and high frequency at the same time in the laboratory.

Motivated by the significance of dynamically assisted Sauter-Schwinger effect, in this paper,
we propose a new extended Sauter-like field model by the combinatorial potentials consisting
of a deep slow-varying potential and some shallow fast-varying potentials, where the frequency
of the fast-varying potential is odd times that of the slow-varying potential. In addition, in our
model, with the increase of the frequency, the intensity of the fast-varying potential decreases
continuously. On the one hand, when the number of combined potentials is increased, the time
shape of superimposed potentials is asymptotic to that of a rectangular wave.

We mainly study the influence of the dynamically assisted Sauter-Schwinger effect on the pair
creation process in different parameter regions by using the CQFT. The enhancement of pair production is found to be significant for some frequency regions and superimposed potentials. The
effective interaction time of bound states is obviously enhanced in some parameter regions. In our
previous work, it was concluded that at low frequencies the effective interaction time of bound
states has a great influence on the number of electrons~\cite{Wangli2019}. Therefore, the proposed model is efficient to the pair production. On the other hand, beside the dynamically assisted Sauter-Schwinger effect, due to the complexity of combined potentials, the pair production process involves the competition of various mechanisms. Our research indicate that their respective contributions to the pair production process are different in different parameter regions. Some new results are found for the pair particles spatial density or/and momentum distribution. Especially an interesting quasi-monoenergetic positron source creation is found when the alternation potential is applied insteading of pure potential well or barrier.

This paper is organized as follows. In Sec. II, we describe multiple alternating potential models. In Sec.III, we discuss and analyze numerical results of the final number, momentum spectrum, and spatial density of electrons at different frequency parameters and potential models. Bound states of the potential for different combinations are analyzed. The enhancement of the dynamically assisted effect under two alternating potentials is also analyzed in detail. In Sec.IV, we summarize our work. The outline of computational quantum field theory is described in Appendix A.

\section{Multiple alternating potential models}\label{section2}
We consider multiple alternating potentials described by the shape function,
\begin{equation}\label{Eq Well}
\begin{aligned}
V(z,t)=V_{n}(t)f(t)S(z),
\end{aligned}
\end{equation}
where $S(z)=\{\tanh[(z-D/2)/W]-\tanh[(z+D/2)/W]\}/2$, $D$ is the width of the potential, $W$ is the width of the potential edge, corresponding to the intensity of the electric field.
The corresponding electric field is
\begin{equation}\label{Eq field}
\begin{aligned}
E(z,t)=\frac{V_{n}(t)}{2W}f(t)\left[\rm{sech}^2\left(\frac{\emph{z}-\emph{D}/2}{\emph{W}}\right)-\rm{sech}^2\left(\frac{\emph{z}+\emph{D}/2}{\emph{W}}\right)\right].
\end{aligned}
\end{equation}
We set $W=0.3\lambda_{\rm{C}}$ and $D=10\lambda_{\rm{C}}$, where $\lambda_{\rm{C}}=1/c$ is the Compton wavelength of electron, $c=1/\alpha$ is the photon speed, $\alpha$ is the fine-structure constant. Throughout this paper the atomic units (a.u.) $m=\hbar=e=1$ are used.
The time-dependent potential depth is
\begin{equation}\label{Eq Vn}
\begin{aligned}
V_{n}(t)=A_{n}V_{0}\sum_{k=1}^{n}\frac{1}{(2k-1)}\sin[(2k-1)\omega t],
\end{aligned}
\end{equation}
where $A_n$ is the normalization factor, $n$ is the number of superimposed potentials, $V_0$ and $\omega$ are the depth and the frequency of the oscillation potential for $n=1$, respectively. To unify the laser energy, we make the integral $\int_{0}^{T/2} V_{n}(t)dt$ equal to the same constant for different superposition potentials, where $T$ is the oscillating period for $n=1$.
The normalization factor $A_n$ is approximately equal to $1.00$, $0.90$, $0.869$, $0.854$, $0.845$, $0.839$, $\cdots$, $\pi/4$ for $n=1,~2,~3,~4,~5,~6,~\cdots~,~\infty$. With the increase of $n$, the normalization factor $A_n$ decreases from $1$ gradually to $\pi/4$. When $n\rightarrow \infty$, the time shape of the potential depth is rectangle wave.
The opening and closing process of the potential is controlled by
\begin{equation}\label{Eq ft}
\begin{aligned}
f(t)= \left \{
\begin{array}{ll}
\sin(\pi t/2t_0)   & if~ 0<t<t_0,\\
1  & if~ t_0<t<t_0+t_1,\\
\cos[\pi(t-t_0-t_1)/2t_0]  &if~ t_0+t_1<t<2t_0+t_1,
\end{array}
\right.
\end{aligned}
\end{equation}
where $t_0$ is the opening and closing time duration of the field, $t_1$ is the action time of the field.

\begin{figure}
\centering\includegraphics[width=10cm]{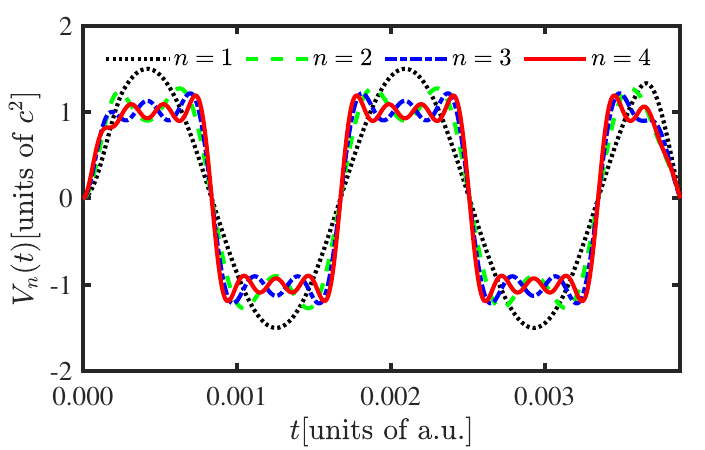}
\caption{\label{fig1} (color online) Depths of multiple alternating potentials as the function of time for $n=1$ (the black dotted curve), $n=2$ (the green dashed curve), $n=3$ (the blue dotted and dashed curve), and $n=4$ (the red solid curve). The other parameters are set to $V_0=1.5c^2$ and $\omega=0.2c^2$.}
\end{figure}
In Fig. \ref{fig1}, depths of multiple alternating potentials as the function of time for $n=1,~2,~3,$ and 4 are presented. Other parameters are set to $V_0=1.5c^2$, and $\omega=0.2c^2$. For $n=1$, the depth amplitude is the maximum.
For $n=2$, superposition potentials consist of a deep slow-varying potential and a shallow fast-varying potential. The high frequency is three times the low frequency. In half an alternating period, there are two peaks.
With the increase of $n$, the depth amplitude gradually decreases, and the depth becomes steeper when alternating, but the alternating period remains the same.
The frequency of each additional potential is an odd multiple of that of the first potential, so the time-dependent function of the potential depth is infinitely close to that of a rectangular wave.

\begin{figure}
\centering\includegraphics[width=10cm]{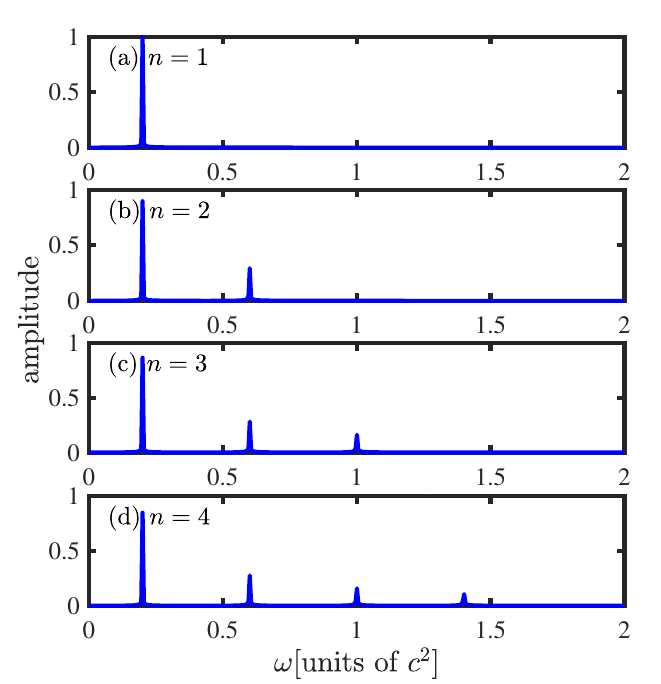}
\caption{\label{fig2} (color online) Frequency spectra of multiple alternating potentials for $n=1$, $2$, $3$, and $4$. Parameters $V_0$ and $\omega$ are the same as in Fig. \ref{fig1}.}
\end{figure}

In Fig. \ref{fig2}, frequency spectra of multiple alternating potentials for $n=1$, $2$, $3$, and $4$ are presented. Parameters $V_0$ and $\omega$ are the same as in Fig. \ref{fig1}. For $n=1$, the frequency is set to $\omega=0.2c^2$.
In Fig. \ref{fig2} (b), the two frequency components are set to $\omega=0.2c^2$ and $3\omega=0.6c^2$, respectively.
The amplitude of the high-frequency component is one-third of that of the low-frequency component. In addition, the amplitude of the low-frequency component is also reduced compared with Fig. \ref{fig2} (a), because the normalization factor $A_2$ is smaller than $1$.
In Figs. \ref{fig2} (c) and (d), high frequency components $5\omega=1.0c^2$ and $7\omega=1.4c^2$ have significantly lower amplitudes.

\section{Numerical results}
In this section, numerical results are presented.
Since this ideal model is complicated at high $n$, we first study simple cases of a single and two alternating potentials in detail before generalizing to the high $n$ case.

\subsection{An alternating potential}

\begin{figure}[htbp]
\centering\includegraphics[width=10cm]{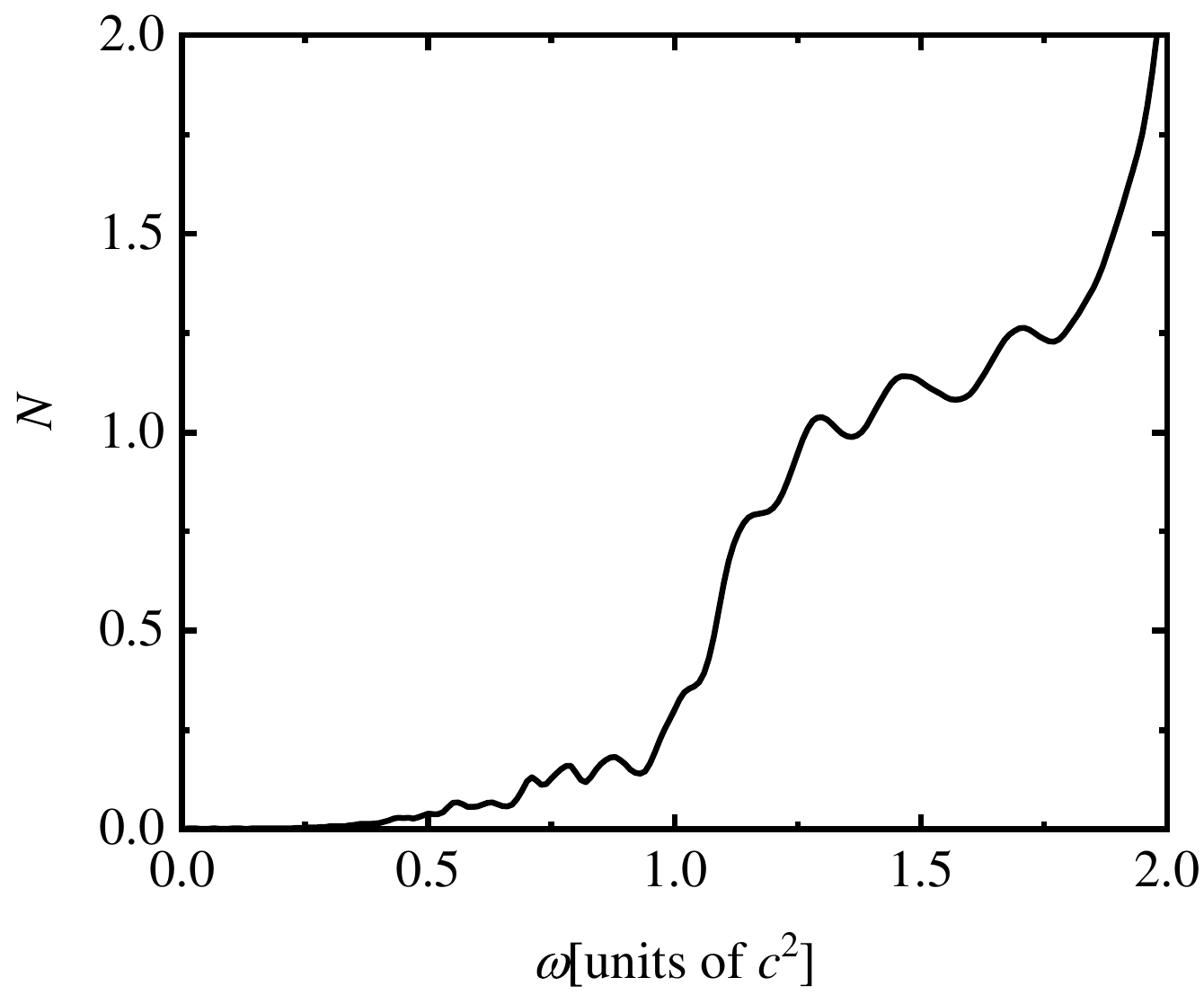}
\caption{\label{fig3} Number of electrons created under an alternating potential as a function of frequency for $V_{1}(t)= V_{0}\sin(\omega t)$. Other parameters are set to $W=0.3/c$, $D=10/c$, $V_0=1.5c^2$, $t_0=5/c^2$a.u., $t_1=20\pi/c^2$a.u., $L=2$.}
\end{figure}
Now, let us discuss the results obtained within an alternating potential.
In Fig. \ref{fig3} we present the final number of created electrons as a function of frequency under an alternating potential. The time-dependent depth of the potential is set to $V_{1}(t)= V_{0}\sin(\omega t)$, where $V_0=1.5c^2$ is the  amplitude. Other parameters are set to $W=0.3/c$, $D=10/c$, $t_0=5/c^2$ a.u., $t_1=20\pi/c^2$ a.u., $L=2$. Pair production is a competing process of several mechanisms.
In the low-frequency region, electron-positron pairs are mainly produced by Schwinger mechanism. Few electrons are produced because of subcritical fields.
When the frequency reaches $2.00c^2$ or $1.00c^2$, the number of electrons increases significantly due to the dominance of one-photon or two-photon process.
There are also a small increase in the number when the frequency is set to $0.4c^2$, $0.5c^2$ and $0.67c^2$, corresponding to five-photon, four-photon and three-photon process, respectively.
Different mechanisms have various effects on the generated electrons.
We track the momentum spectrum and the spacial density of generated electrons at several frequency parameters.
This intuitive description of the pair generation process helps to understand the different production mechanisms.

\begin{figure}[htbp]
\centering\includegraphics[width=15cm]{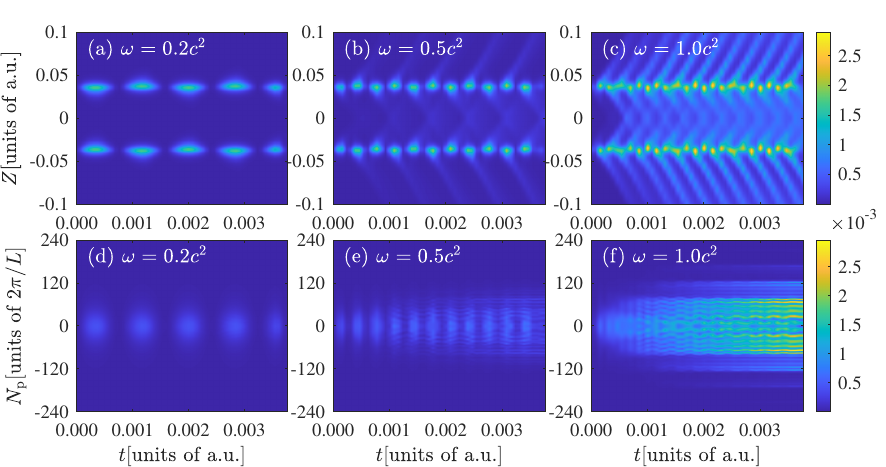}
\caption{\label{fig4} (color online) Time evolutions of the spacial density (top row) and momentum spectrum (bottom row) of electrons created under an alternating potential for $\omega=0.2c^2$, $\omega=0.5c^2$ and $\omega=1.0c^2$. Other parameters are the same as in Fig. \ref{fig3}.}
\end{figure}
In Fig. \ref{fig4}, time evolutions of the spacial density (top row) and the momentum spectrum (bottom row) of created electrons under an alternating potential for different frequencies is presented. Other parameters are the same as in Fig. \ref{fig3}.
When $\omega=0.2c^2$, the pair creation is dominated by Schwinger mechanism. In Fig. \ref{fig4} (a), electrons mainly gather on both sides of the potential, i.e., the location of large electric field strength. The pair creation rate under Schwinger mechanism is proportional to the electric field intensity.
The spatial density of electrons oscillates periodically with time. This is because the electric field strength increases or decreases periodically with time. In Fig. \ref{fig4} (d), the time evolution of the momentum spectrum also has a significant time effect. In addition, the momentum of created electrons is too small to overcome the electric field. Hence, electrons are mainly distributed in the low momentum region.
When $\omega=0.5c^2$, the momentum spectrum is more widely distributed due to four-photon process. Some electrons move inside or outside the potential, which is related to the reversal of the electric field direction.
When $\omega=1.0c^2$, multi-photon processes are enhanced, especially two-photon process.
The momentum spectrum broadens to the high frequency region, resulting in a large number of electrons escaping from the potential in Fig. \ref{fig4} (c). The time effect of the momentum spectrum in Fig. \ref{fig4} (f) gradually disappears.

\begin{figure}[htbp]
\centering\includegraphics[width=14cm]{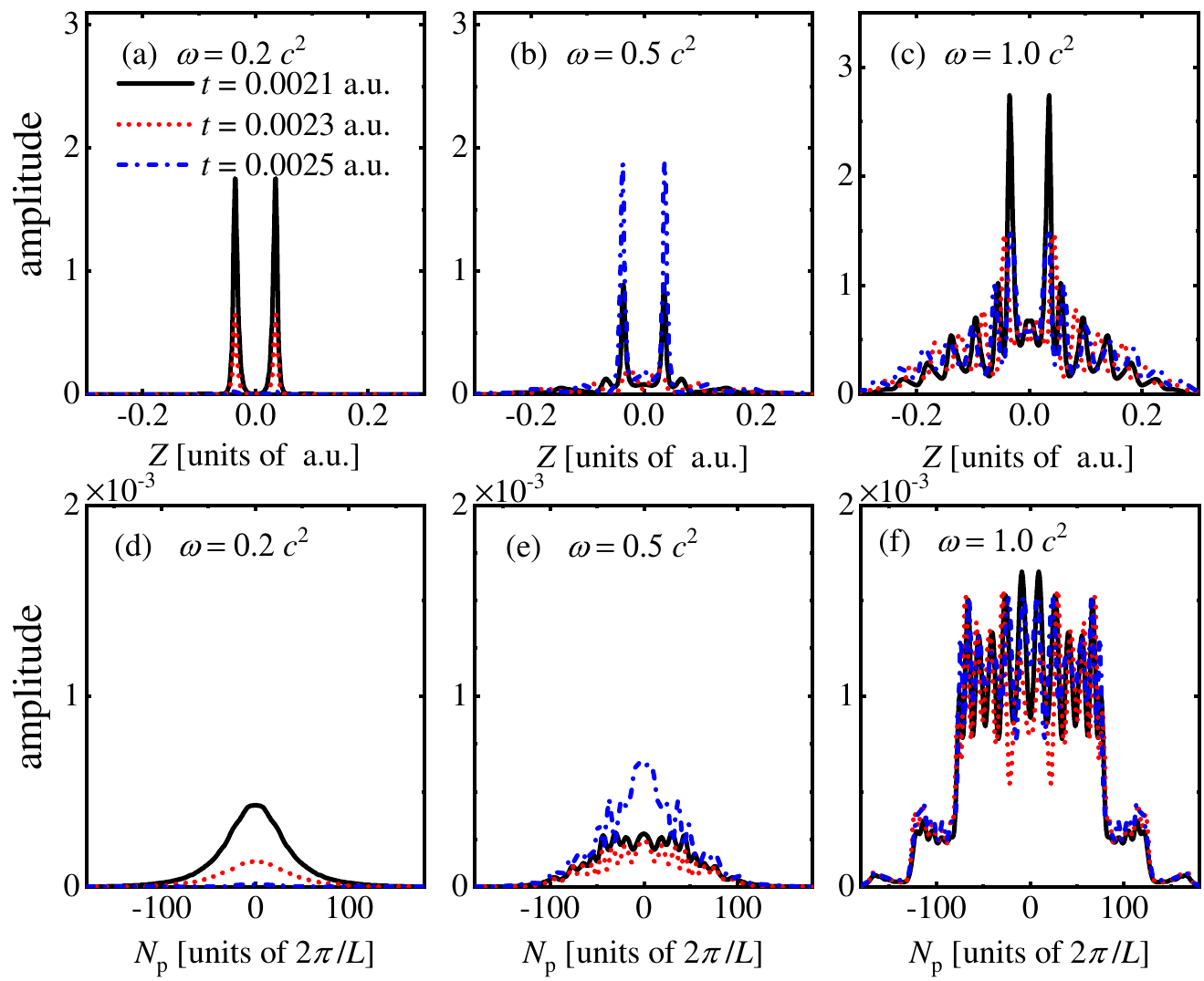}
\caption{\label{fig5} (color online) Spacial density (up row) and momentum spectrum (bottom row) distributions of electrons created under an alternating potential for $t=0.0021$ a.u. (black solid curves), $0.0023$ a.u. (red dotted curves), and $0.0025$ a.u. (blue dot-dash curves). Frequency parameter is the same as in Fig. \ref{fig4}. Other parameters are the same as in Fig. \ref{fig3}.}
\end{figure}
In Fig. \ref{fig5}, the spacial density and momentum spectrum distributions of created electrons for $t=0.0021$ a.u., $0.0023$ a.u. and $0.0025$ a.u. are presented. The frequency is the same as in Fig. \ref{fig4}. Other parameters are the same as in Fig. \ref{fig3}. Negative momentum means electrons moving in the opposite direction.
In Figs. \ref{fig5} (a) and (d), both the spatial density and the momentum spectrum change dramatically with time. Under the domination of Schwinger mechanism, the momentum spectrum amplitude decays exponentially with increasing momentum.
In Figs. \ref{fig5} (c) and (f), the time dependence of the electron spatial density and momentum spectrum distribution is weak. Under the action of multi-photon processes, the momentum is distributed in steps, and each step has multiple peaks. These peaks are associated with discrete bound state energy levels in the potential \cite{Tang2013}.
When $\omega =0.5c^2$, in Fig. \ref{fig5} (b) and (e), the amplitudes and profiles of the electron spatial density and momentum spectrum are comparable to those of $\omega =0.2c^2$, reflecting the dominance of the Schwinger mechanism. Multiple peaks of the momentum spectrum also reflect the role of multiphoton processes.
The strong time effect of the momentum spectrum of the generated electrons and the exponential decay distribution are the characteristics of the Schwinger mechanism.
Overall, the strong time effect and the exponentially decaying distribution of the momentum spectrum are characteristics of Schwinger mechanism. Multiphoton processes are characterized by multiple peaks in the momentum spectrum.

In the Dirac vacuum, particles in the negative energy continuum can leap directly to the positive energy continuum through Schwinger mechanism or multiphoton processes thereby generating  electron-positron pairs, or it is possible that the particle first leaps into the energy gap and then absorbs multiple photons. In addition, discrete bound states in the potential also provide a platform for particles, increasing the chance of leap. The position of bound states in the energy gap has a significant effect on the energy spectrum of created electrons.
Next, we analyze the energy spectrum of electrons produced under an alternating potential, an oscillating potential well, and an oscillating potential barrier in conjunction with the instantaneous bound states.

\begin{figure}[htbp]
\centering\includegraphics[width=15cm]{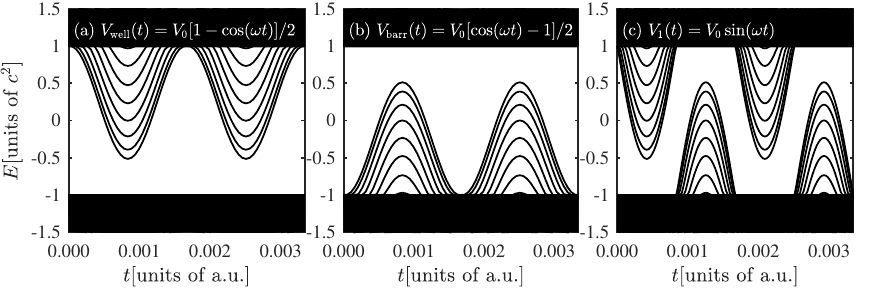}
\caption{\label{fig6} Instantaneous eigenvalues of potentials over time for (a) the oscillating potential well $V_{\textrm{well}}(t)= V_{0}[1-\cos(\omega t)]/2$, (b) the oscillating potential barrier $V_{\textrm{barr}}(t)= V_{0}[\cos(\omega t)-1]/2$, and (c) the alternating potential $V_1(t)= V_{0}\sin(\omega t)$, where $\omega=0.2c^2$. Other parameters are the same as in Fig. \ref{fig3}.}
\end{figure}

In Fig. \ref{fig6}, we display the instantaneous energy eigenvalues of the alternating potential, the oscillating potential well, and the oscillating potential barrier over time, by diagonalizing the Dirac
Hamiltonian at every instant of time. The frequency is set to $\omega=0.2c^2$. Other parameters are the same as in Fig. \ref{fig3}.
These three potentials have the same number of bound states, which is due to the equal maximum potential depth.
For the potential well, discrete bound states come from the positive energy continuum in Fig. \ref{fig6} (a). The potential well is attractive for the electron.
Conversely, for the barrier, bound states emerge from the negative energy continuum. The potential barrier is attractive for the positron.
For an alternating potential, bound states alternately emerge from the positive or the negative energy continuum. Therefore, in Fig. \ref{fig4}, electrons generated at different times alternately enter and exit the potential.
The energy and momentum spectra of the electrons produced by different potentials are different because of the different positions of bound states.

\begin{figure}[htbp]
\centering\includegraphics[width=10cm]{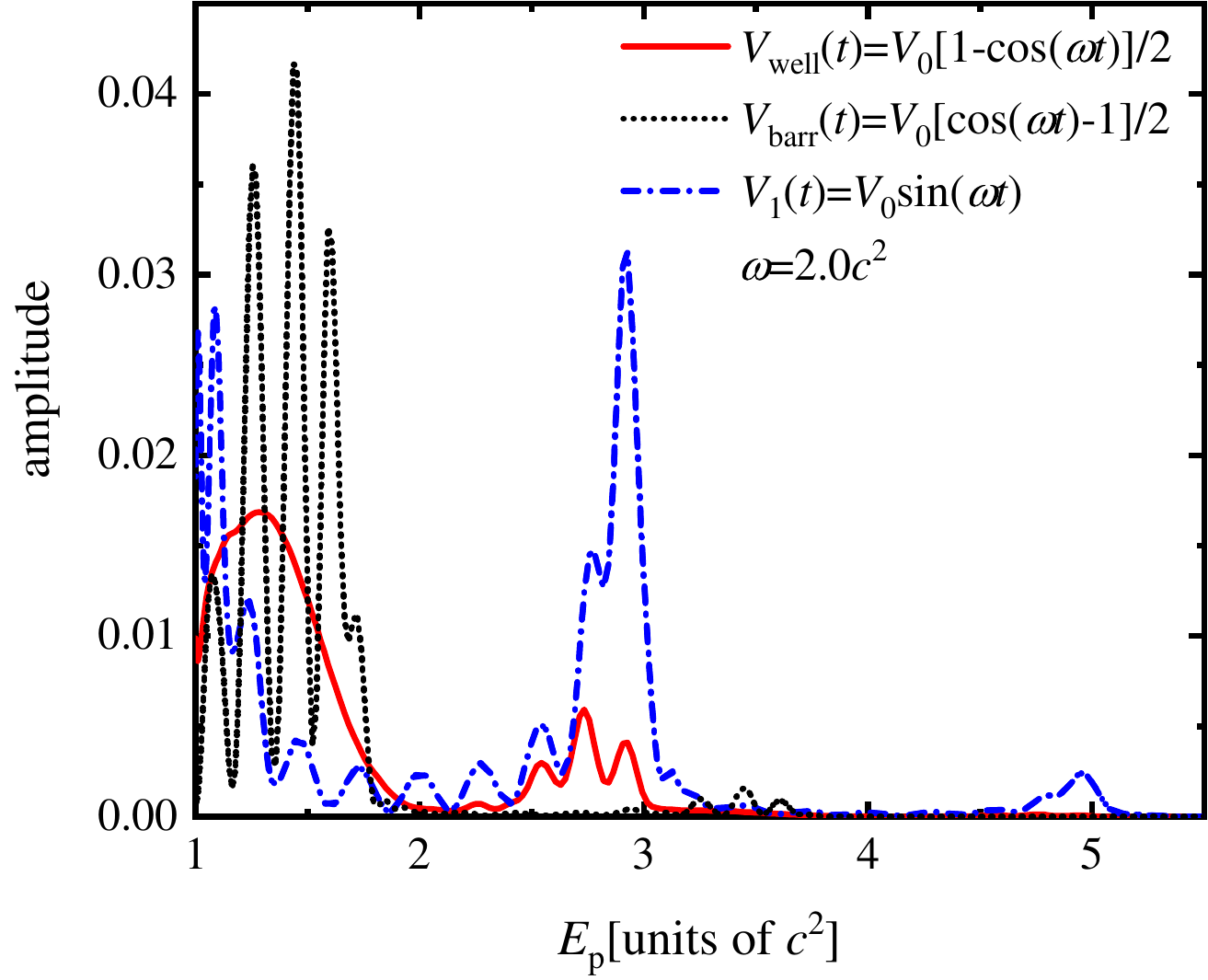}
\caption{\label{fig7} (color online) Energy spectra of electrons created under the oscillating potential well (the red solid curve), the oscillating potential barrier (the black dotted curve), and the alternating potential (the blue dot-dash curve). The frequency is set to $\omega=2.0c^2$. Other parameters are the same as in Fig. \ref{fig3}.}
\end{figure}
In Fig. \ref{fig7}, energy spectra of electrons created under the oscillating potential well, the oscillating potential barrier and the alternating potential are presented. The frequency is set to $\omega=2.0c^2$.
For the potential well, the curve is smooth in the region of $1.0c^2<E_{\mathrm{p}}<2.0c^2$ and has multiple peaks in the region of $2.0c^2<E_{\mathrm{p}}<3.0c^2$.
For the potential barrier, the peaks appear in the region of $1.0c^2<E_{\mathrm{p}}<2.0c^2$. For alternating potentials, the peaks cover a wider region of $1.0c^2<E_{\mathrm{p}}<3.0c^2$.
When $E_{\mathrm{p}}=5.0c^2$, a sub peak appears.
Compared to the oscillating potential well and barrier, the electrons produced in the alternating potential have higher energy and are more monoenergetic. The quasimonoenergetic positron source would be useful in the electron-positron colliding experiments~\cite{Tang2021,Cabibbo1961,Myers1990,Tang2019}.

When $\omega=2.0c^2$, electron-positron pairs can mainly be created by the Schwinger mechanism, the two-photon process, or the combination of the Schwinger mechanism and the one-photon process, where the Schwinger mechanism is relatively weak. Since the probability of two-photon process is an order of magnitude smaller than that of one-photon process, it is the combination of the Schwinger mechanism and the one-photon process that dominates the pair production. Hence, the energy of created electrons is mainly distributed in the region of $1.0c^2<E_{\mathrm{p}}<3.0c^2$.
The particle in the negative energy continuum first tunnels into the energy gap and then absorbs a photon to jump to the positive energy continuum. With the increase of the energy level in the energy gap, the tunneling rate decays, which explains why the red solid curve decays in the region of $1.0c^2<E_{\mathrm{p}}<2.0c^2$ in Fig. \ref{fig7}.
If there are bound states in the energy gap, there is a high probability that a particle in the negative energy continuum will tunnel into the bound state. The production rate is proportional to the number of bound states and the interaction time.
Since bound states of the potential well are mostly located in the positive energy gap, the peak mainly appears in the region of $2.0c^2<E_{\mathrm{p}}<3.0c^2$. The difference in peak value is related to the interaction time of bound states.
Similarly, bound states of the potential barrier are mainly located in the negative energy gap, resulting in the peak distribution mainly in the region from from $1.0c^2$ to $2.0c^2$. For the alternating potential, peaks are distributed from $1.0c^2$ to $3.0c^2$ because bound states alternate between the positive and the negative energy gap. Peaks in the region of $3.0c^2<E_{\mathrm{p}}<5.0c^2$
correspond to the combination of the Schwinger mechanism and the two-photon process. It is clear that the peak value drops by an order of magnitude.

Note that an interesting finding is for the particles of pair produced in the alternating potential, see the bottom of Fig. \ref{fig5}, which has a quasi-monoenergetic structure compared to cases of the oscillating potential well or/and potential barrier in Fig. \ref{fig7}.

\subsection{Two alternating potentials}
Next, we investigate the enhancement of pair creation in two alternating potentials. The potential depth as a function of time is set to $V_{2}(t)=0.9V_{0}\sin(\omega t)+0.3V_{0}\sin(3\omega t)$, where $V_0=1.5c^2$. The effect of frequency variation on the pair production process is studied. Numerical results of the electron number, momentum spectrum and spacial density distribution at different frequencies are shown below.

\begin{figure}[htbp]
\centering\includegraphics[width=10cm]{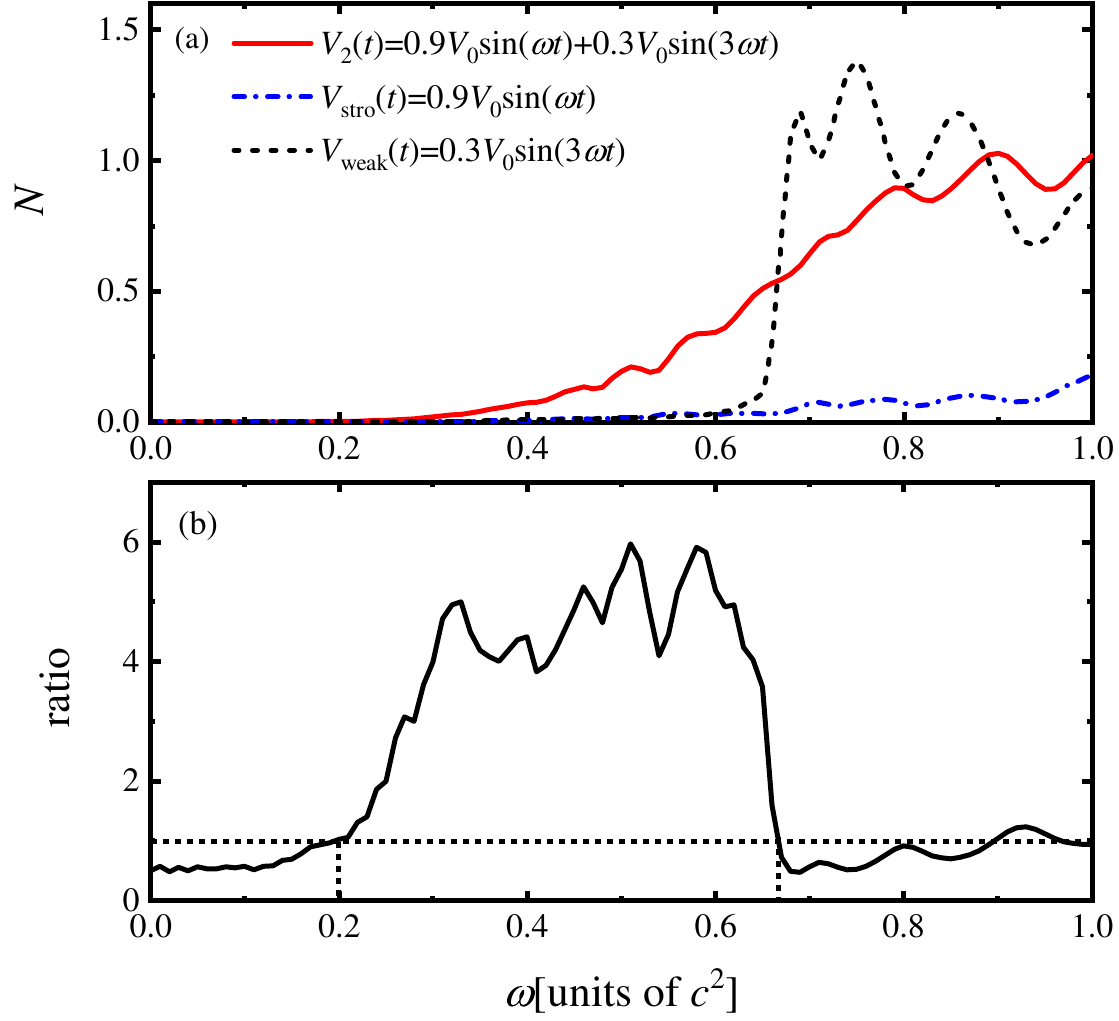}
\caption{\label{fig8} (color online) (a) Number of electrons as a function of frequency for $V_{2}(t)=0.9V_{0}\sin(\omega t)+0.3V_{0}\sin(3\omega t)$ (the red solid curve), $V_{\mathrm{stro}}(t)=0.9V_{0}\sin(\omega t)$ (the black dashed curve) and $V_{\mathrm{weak}}(t)=0.3V_{0}\sin(3\omega t)$. (b) The ratio of the number of electrons produced in two alternating potentials to the sum of the number of electrons produced in two single alternating potentials. The black dashed line indicates that the ratio is equal to $1$. Other parameters are the same as in Fig. \ref{fig3}.}
\end{figure}
In Fig. \ref{fig8} (a), number of electrons as a function of frequency for $V_{2}(t)=0.9V_{0}\sin(\omega t)+0.3V_{0}\sin(3\omega t)$, $V_{\mathrm{stro}}(t)=0.9V_{0}\sin(\omega t)$ and $V_{\mathrm{weak}}(t)=0.3V_{0}\sin(3\omega t)$ is presented. The frequency is set to vary from $0$ to $1.0c^2$. Strictly physically speaking, the maximum frequency $3\omega$ should not exceed $2.0c^2$, i.e., $\omega<0.67c^2$.
Theoretically, it can be extended to the supercritical region of $3\omega>2.0c^2$, which can yield some unexpected gains.
When $\omega<0.67c^2$, the number of electrons produced under both the fast-varying and the slow-varying potential is very small. However, for combined potentials, the electron number increases significantly. When $0.67c^2<\omega<1.0c^2$, the number of electrons remains small for the slow-varying potential, which is consistent with Fig. \ref{fig3}. For the fast-varying potential, the number of electrons first rises rapidly and then slowly oscillates down.
This is because the actual frequency $3\omega$ exceeds the critical value of $2.0c^2$, leading to the enhancement of multiphoton processes, especially the one-photon process. However, continuing to increase the frequency will have a suppressive effect on the pair production \cite{Jiang2013}.
When the maximum frequency exceeds the critical frequency, there is no advantage of combined potentials compared to the individual potential.

In Fig. \ref{fig8} (b), the ratio of the number of electrons produced in two alternating potentials to the sum of the number of electrons produced in two single alternating potentials is presented. The black dotted line is for $R=1$.
In the range of $0<\omega<0.2c^2$, the ratio is less than $1$. The high frequency component ranges from $0$ to $0.6c^2$, with the Schwinger mechanism still playing a dominant role. The combined potential reduces the peak intensity of the electric field, resulting in a decrease in the number of electrons.
When $0.2c^2<\omega<0.67c^2$, the number of electrons can be increased by six times. The high frequency component ranges from $0.6c^2$ to $2.0c^2$. The combination of low-frequency strong fields and high-frequency weak fields enhances the dynamically assisted Sauter-Schwinger effect.
When $\omega>0.67c^2$, the high frequency component $3\omega$ enters the supercritical region, leading to a decrease in the yield of the pair.

The Keldysh parameter $\gamma=m c\omega/|e E|$ is an important parameter for the study of the pair creation from vacuum under the alternating field \cite{Brezin1970,Keldysh1965}, where $E$ is the field strength, $\omega$ is the frequency, $m$ and $e$ are the mass and the charge of the electron, respectively. When $\gamma\ll1$, electron-positron pairs are mainly created by Schwinger mechanism. When $\gamma\gg1$, multiphoton processes dominate the pair generation. The Schwinger mechanism and multiphoton processes work together when the Keldysh parameter tends to $1$.
The dynamically assisted Sauter-Schwinger effect is enhanced for the combination of a strong field with $\gamma\ll1$ and a weak field with$\gamma\gg1$.
In the following we discuss physical mechanisms revealed by this important parameter in a few specific examples.

Firstly, for the chosen frequency $\omega=0.1c^2$, the Keldysh parameter is approximately $\gamma_{\rm{stro}}\simeq0.044$ for the strong field, indicating the dominance of the Schwinger mechanism.
For the weak field, the Keldysh parameter is set to $\gamma_{\rm{weak}}=0.40$, indicating the combined mechanism of Schwinger mechanism and multiphoton processes.
Since both $\gamma_{\rm{stro}}$ and $\gamma_{\rm{weak}}$ are less than $1$, the dynamically assisted Sauter-Schwinger effect is very weak. For small $\gamma$, we can analyze the pair generation process in conjunction with the time evolution of the instantaneous bound state.

\begin{figure}[htbp]
\centering\includegraphics[width=15cm]{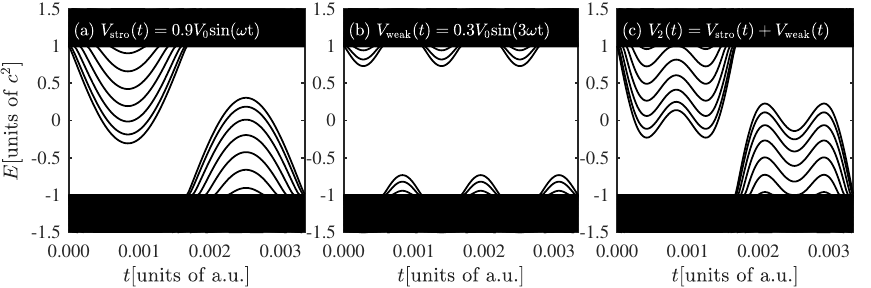}
\caption{\label{fig9} Instantaneous eigenvalues of potentials over time for (a) the slow-varying potential $V_{\textrm{stro}}(t)= 0.9V_{0}\sin(\omega t)$, (b) the fast-varying potential $V_{\textrm{weak}}(t)= 0.3V_{0}\sin(3\omega t)$, and (c) two alternating potentials $V_2(t)= 0.9V_{0}\sin(\omega t)+0.3V_{0}\sin(3\omega t)$, where $\omega=0.1c^2$. Other parameters are the same as in Fig. \ref{fig3}.}
\end{figure}
In Fig. \ref{fig9}, we display the instantaneous energy eigenvalues of the slow-varying potential, the fast-varying potential, and two alternating potentials over time. The frequency is set to $\omega=0.1c^2$.
The number of bound states is proportional to the depth of the potential. Therefore, in Figs. \ref{fig9} (a) and (b), there are fewer bound states for the weak fast potential compared to the strong slow potential.
In Fig. \ref{fig9} (c), the alternating period of bound states of combined potentials is the same as that of the strong slow-varying potential. The difference is that two peaks appear within half an alternating period of combined potentials, which is caused by the high frequency component $3\omega$.
The duration of bound states is prolonged by the superposition of potentials. In addition, the energy levels of bound states are reduced, which also reduces the rate of pair generation.

\begin{figure}[htbp]
\centering\includegraphics[width=15cm]{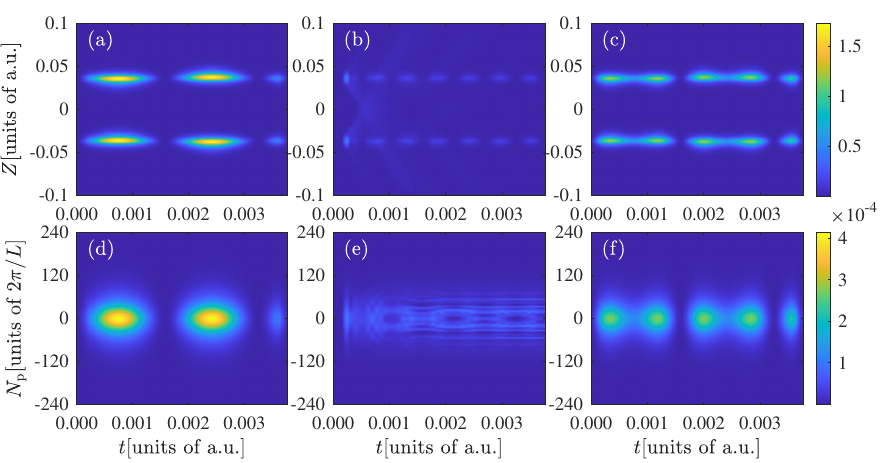}
\caption{\label{fig10} (color online) Time evolutions of the spacial density (top row) and momentum spectrum (bottom row) of electrons created under the slow-varying potential $V_{\textrm{stro}}(t)= 0.9V_{0}\sin(\omega t)$ (left column), the fast-varying potential $V_{\textrm{weak}}(t)= 0.3V_{0}\sin(3\omega t)$ (middle column), and two alternating potentials $V_2(t)= 0.9V_{0}\sin(\omega t)+0.3V_{0}\sin(3\omega t)$ (right column), where $\omega=0.1c^2$. Other parameters are the same as in Fig. \ref{fig3}.}
\end{figure}
In Fig. \ref{fig10}, we display the time evolutions of the spacial density (top row) and the momentum spectrum (bottom row) of created electrons under the slow-varying potential, the fast-varying potential, and two alternating potentials. The two frequency parameters are set to $\omega=0.1c^2$ and $3\omega=0.3c^2$.
Other parameters are the same as in Fig. \ref{fig3}. In Figs. \ref{fig10} (a), (b) and (c), electrons are mainly distributed at edges of the potential.
For the slow-varying potential, the momentum spectrum and spacial density significantly vary over time. The Schwinger mechanism dominates the process of pair generation. In Fig. \ref{fig10} (e), the time dependence and multiple streaks of the momentum spectrum together reflect a combined mechanism of the Schwinger mechanism and multiphoton processes.
For two alternating potentials, the Schwinger mechanism also dominates the pair generation process. Time evolution periods of the number density and momentum spectrum are equal to that of the slow-varying potential. There are two peaks in each cycle under two alternating potentials due to the superposition of potentials.
When pair creation process is dominated by Schwinger mechanism,
time evolution periods of the electron density and momentum spectrum are two times that of the instantaneous bound states. Amplitudes of spacial density and momentum spectrum are proportional to the number of bound states.

\begin{figure}[htbp]
\centering\includegraphics[width=15cm]{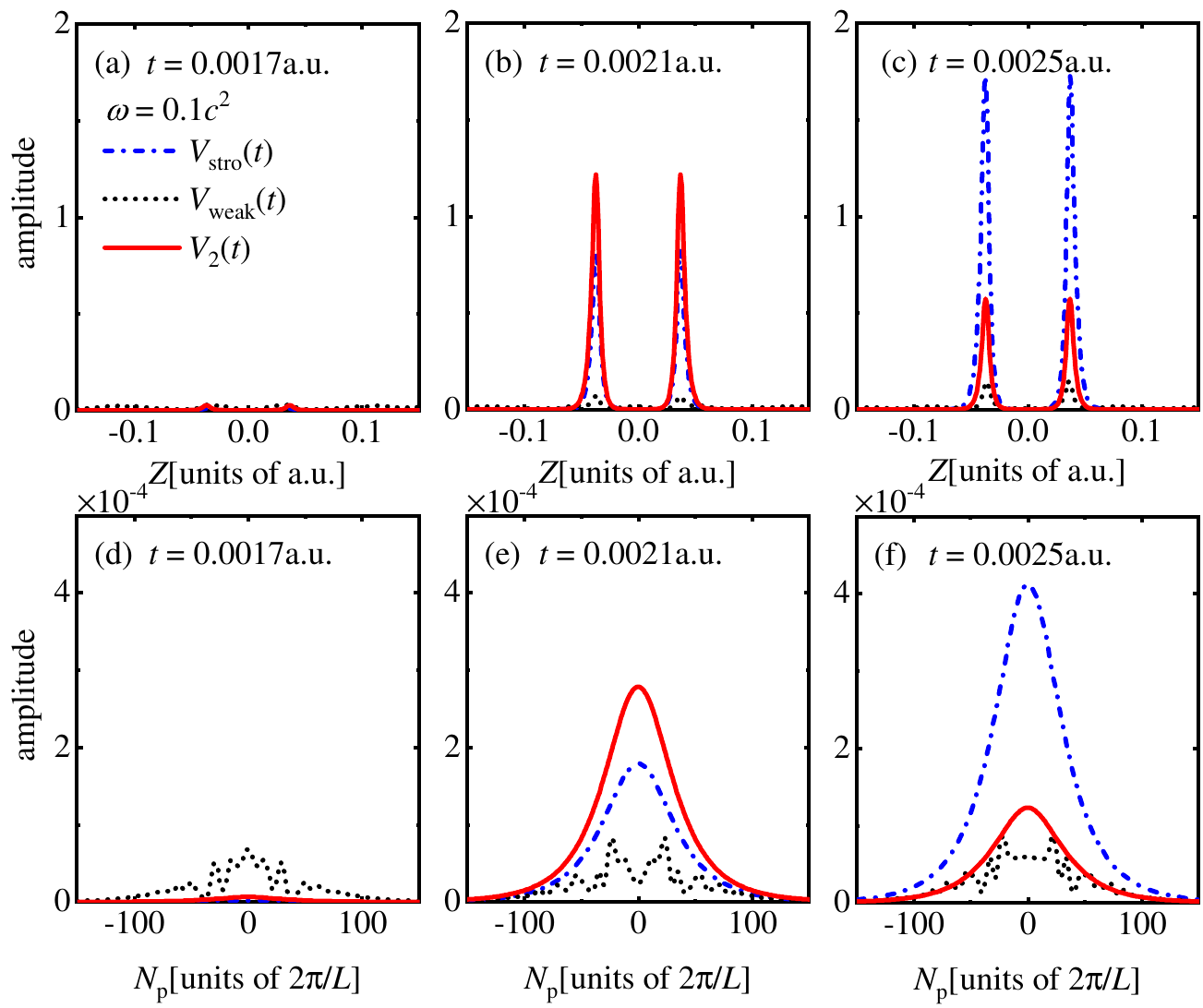}
\caption{\label{fig11} (color online) Spacial density (up row) and momentum spectrum (bottom row) distributions in Fig. \ref{fig10} for $t=0.0017$ a.u. (left column), $t=0.0021$ a.u. (middle column), and $t=0.0025$ a.u. (right column). The frequency is set to $\omega=0.1c^2$. Other parameters are the same as in Fig. \ref{fig3}.}
\end{figure}
In Fig. \ref{fig11}, spacial density (up row) and momentum spectrum (bottom row) distributions in Fig. \ref{fig10} at different times are presented.
When $t=0.0017$ a.u., amplitudes of spacial density and momentum spectrum are very small for the slow-varying potential and two alternating potentials.
When $t=0.0021$ a.u., the amplitude reaches the maximum for two alternating potentials. When $t=0.0025$ a.u., the amplitude reaches the maximum for the slow-varying potential. The maximum amplitude of two alternating potentials is smaller than that of the slow-varying potential. This is because the combination of potentials leads to a reduction in the number of bound states. In addition, for the  slow-varying potential and two alternating potentials, momentum spectra exhibits a smooth exponential exponentially decays with increasing momentum. For the fast-varying potential, there are many peaks under the exponential envelope, which is the result of the coupling of Schwinger mechanism and multiphoton processes.

Secondly, when the frequency is $0.5c^2$, the Keldysh parameter is approximately $\gamma_{\rm{stro}}\simeq0.22$ for the strong field, $\gamma_{\rm{weak}}=2.0$ for the weak field.
For combined fields, the dynamically assisted Sauter-Schwinger effect has an enhancement on the generation of electron-positron pairs.

\begin{figure}[htbp]
\centering\includegraphics[width=15cm]{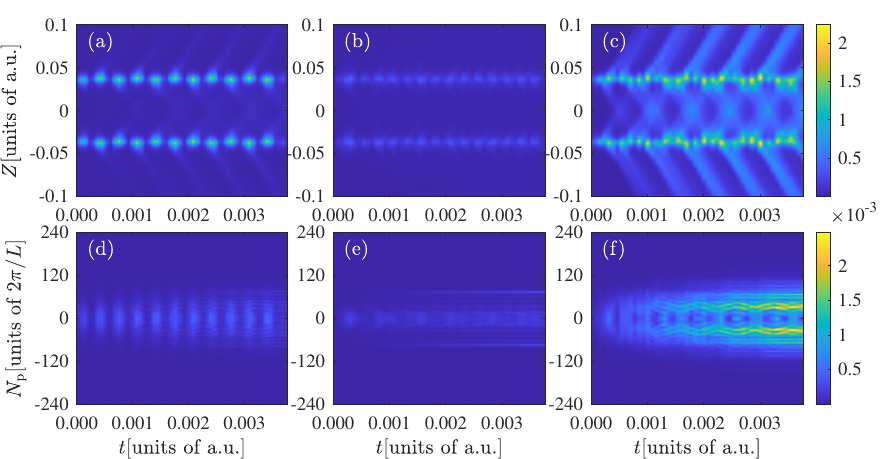}
\caption{\label{fig12} (color online) Time evolutions of the spacial density (top row) and momentum spectrum (bottom row) of electrons created under the slow-varying potential $V_{\textrm{stro}}(t)= 0.9V_{0}\sin(\omega t)$ (left column), the fast-varying potential $V_{\textrm{weak}}(t)= 0.3V_{0}\sin(3\omega t)$ (middle column), and two alternating potentials $V_2(t)= 0.9V_{0}\sin(\omega t)+0.3V_{0}\sin(3\omega t)$ (right column), where $\omega=0.5c^2$. Other parameters are the same as in Fig. \ref{fig3}.}
\end{figure}
In Fig. \ref{fig12}, we display time evolutions of the spacial density (top row) and momentum spectrum (bottom row) of electrons created under the slow-varying potential (left column), the fast-varying potential (middle column), and two alternating potentials. The frequencies are set to $\omega=0.5c^2$ and $3\omega=1.5c^2$. Other parameters are the same as in Fig. \ref{fig3}. Similar to the case of $\omega=0.5c^2$ in Fig. \ref{fig4}, both Schwinger mechanism and multiphoton processes affect the pair generation under the slow-varying potential. For the fast-varying potential, the time evolution of the momentum spectrum reflects the dominance of multiphoton processes. For two alternating potentials, the combination of strong slow fields and weak fast fields promotes the dynamically assisted effect. In Fig. \ref{fig12} (c), many electrons escape from the combined potentials, resulting in an increase in the number of electrons. In Fig. \ref{fig12} (f), a large number of electrons are distributed in the high energy region. The time effect of momentum spectrum of electrons is weakened.

\begin{figure}[htbp]
\centering\includegraphics[width=15cm]{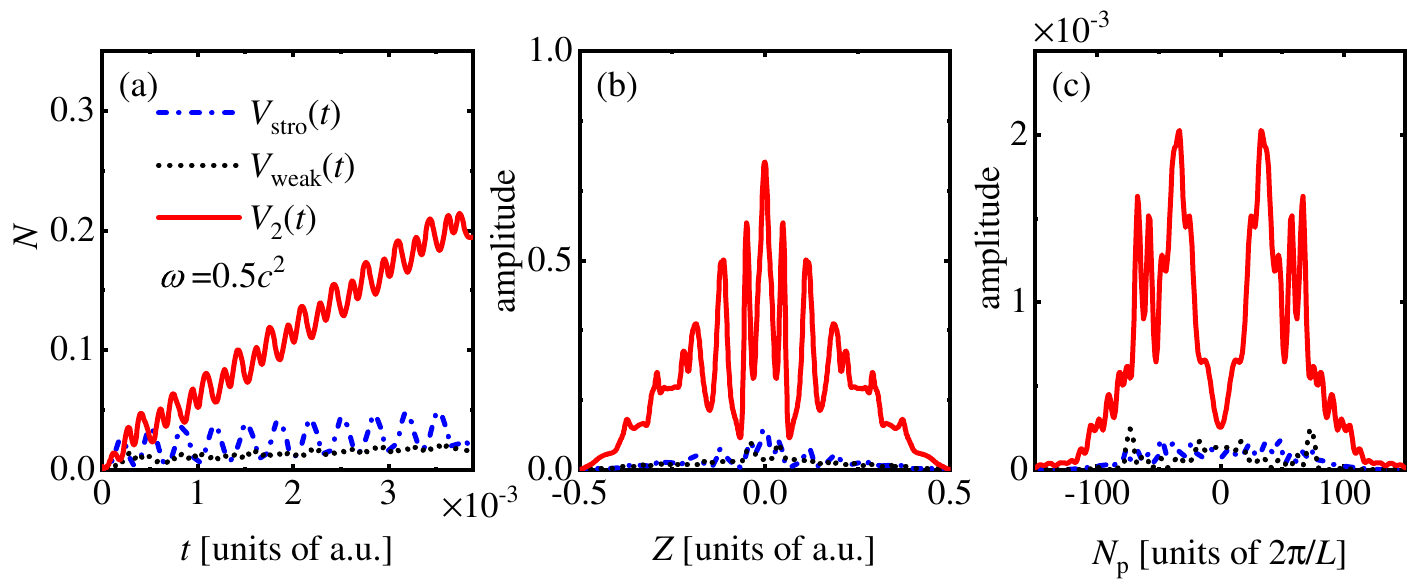}
\caption{\label{fig13} (color online) (a) Number of electrons as a function of time. (b) The spacial density and (c) momentum spectra of electrons at the last moment in Fig. \ref{fig12}. The blue dot-dash curve is for the slow-varying potential $V_{\textrm{stro}}(t)= 0.9V_{0}\sin(\omega t)$, the black dashed curve is for the fast-varying potential $V_{\textrm{weak}}(t)= 0.3V_{0}\sin(3\omega t)$, and the red solid curve is for two alternating potentials $V_2(t)= 0.9V_{0}\sin(\omega t)+0.3V_{0}\sin(3\omega t)$.
The frequency is set to $\omega=0.5c^2$. Other parameters are the same as in Fig. \ref{fig3}.}
\end{figure}
In Fig. \ref{fig13} (a), number of electrons as a function of time is presented. The frequencies are set to $\omega=0.5c^2$ and $3\omega=1.5c^2$. For the slow-varying potential, the number of electrons slowly oscillates over time. The final number of electrons is about $0.029$. For the fast-varying potential, although the potential oscillates quickly, the final number is as low as $0.037$ due to the weak field strength. For two alternating potentials, the number of electrons increases rapidly over time, eventually reaching $0.345$. The final number is about five times the sum of that number of electrons produced under the fast- and the slow-varying potentials.
There are two peaks in each oscillation cycle, which is associated with the combination of the two potentials.

In Figs. \ref{fig13} (b) and (c), the spacial density and momentum spectra of electrons at the last moment in Fig. \ref{fig12} are presented, respectively. Amplitudes of spacial density and momentum spectra are very small for both slow- and fast- varying potentials. For two alternating potentials, the amplitude is increases significantly.
In Fig. \ref{fig13} (b), each peak represents the generation of an bunch of electrons. With the increase of time, the electron and positron bunches are alternately output. The pumping electron-positron pairs from combined potentials can also provide a source of particles for the electron-positron colliding experiments.
In Fig. \ref{fig13} (c),
the momentum spectrum is mainly distributed in the low momentum region for the slow-varying potential, and in the high frequency region for the fast-varying potential. Combined potentials accumulate advantages of both potentials. These peaks are associated with discrete bound states. The dynamically assisted effect is enhanced by the combination of multiple bound states from the slow-varying potential and high-frequency photons from the fast-varying potential.

Finally, we consider the case where the high frequency component exceeds the critical value of $2.0c^2$. When $\omega=0.75c^2$ and $3\omega=2.25c^2$, the Keldysh parameter is approximately $\gamma_{\rm{stro}}\simeq0.33$ for the strong field, $\gamma_{\rm{weak}}=3.0$ for the weak field. The combination Keldysh is set to $\gamma_{\rm{c}}=1.0$, indicating a strong dynamically assisted effect. However, the final number of electrons is reduced for combined potentials. For this phenomenon, the following spatial density and momentum spectrum distributions will give an explanation.

\begin{figure}[htbp]
\centering\includegraphics[width=15cm]{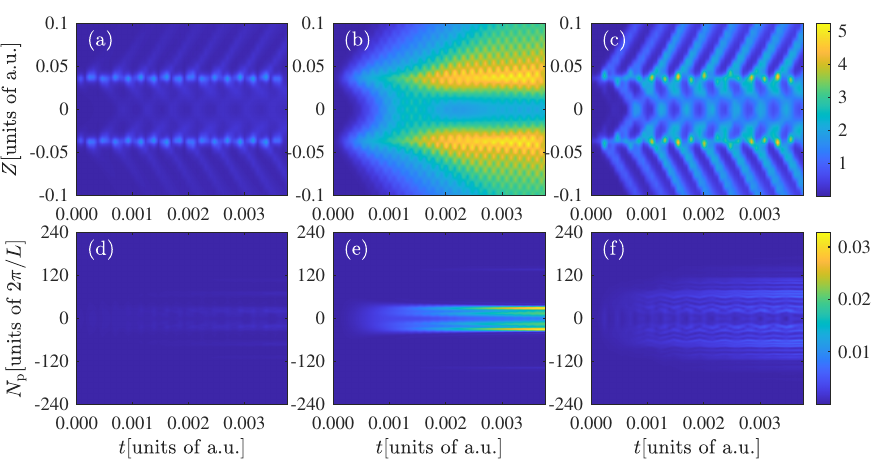}
\caption{\label{fig14} (color online) Time evolutions of the spacial density (top row) and momentum spectrum (bottom row) of electrons created under the slow-varying potential $V_{\textrm{stro}}(t)= 0.9V_{0}\sin(\omega t)$ (left column), the fast-varying potential $V_{\textrm{weak}}(t)= 0.3V_{0}\sin(3\omega t)$ (middle column), and two alternating potentials $V_2(t)= 0.9V_{0}\sin(\omega t)+0.3V_{0}\sin(3\omega t)$ (right column), where $\omega=0.75c^2$. Other parameters are the same as in Fig. \ref{fig3}.}
\end{figure}
In Fig. \ref{fig14}, we display time evolutions of the spacial density (top row) and momentum spectrum (bottom row) of electrons created under the slow-varying potential (left column), the fast-varying potential (middle column), and two alternating potentials. The frequencies are set to $\omega=0.75c^2$ and $3\omega=2.25c^2$. Other parameters are the same as in Fig. \ref{fig3}.
For the slow-varying potential, the three-photon process dominates the pair generation. Many electrons gain enough energy to escape the potential. Compared with the fast-varying potential, amplitudes of spatial density and momentum spectrum are quite small for the slow-varying potential. This is because the one-photon process dominates the pair generation for the fast-varying potential. Furthermore, for the fast-varying potential, although electrons are mainly distributed in the low-momentum region, there are still a large number of electrons distributed outside the potential due to the low electric field strength.
For two alternating potentials, the increase in the number of bound states results in more peaks and a more broader distribution of the momentum spectrum compared to the fast-varying potential. However, the increase in the electric field strength traps some low-energy electrons, resulting in a decrease in the total number of electrons.

\begin{figure}[htbp]
\centering\includegraphics[width=15cm]{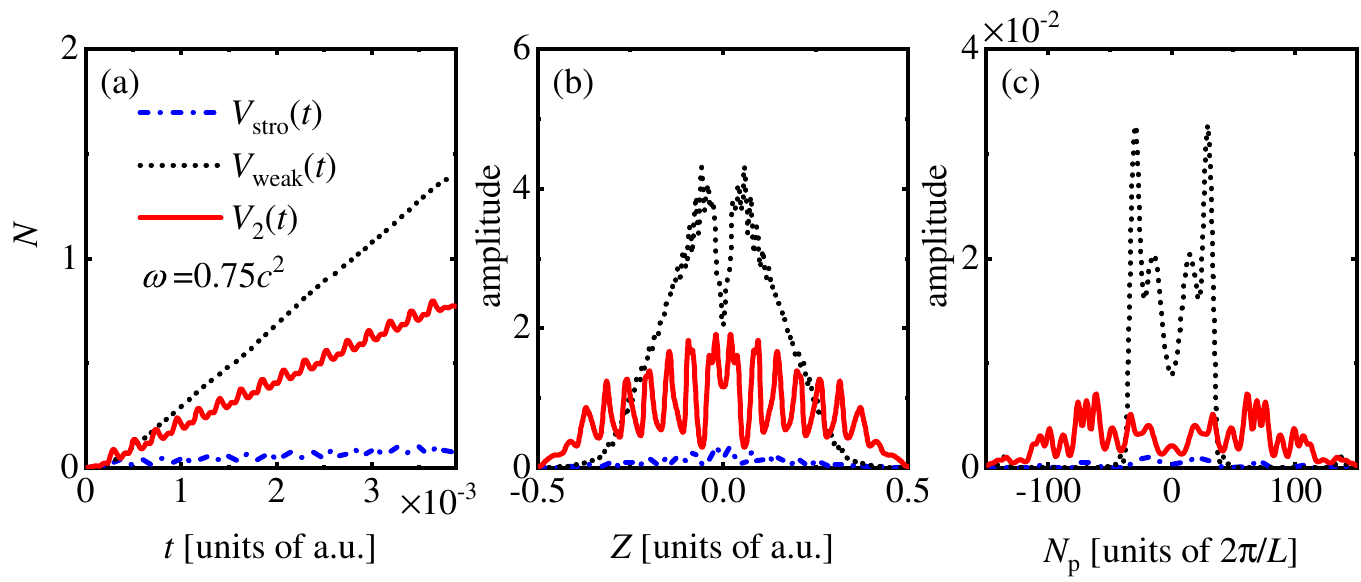}
\caption{\label{fig15}(color online) (a) Number of electrons as a function of time. (b) The spacial density and (c) momentum spectra of electrons at the last moment in Fig. \ref{fig12}. The blue dot-dash curve is for the slow-varying potential $V_{\textrm{stro}}(t)= 0.9V_{0}\sin(\omega t)$, the black dashed curve is for the fast-varying potential $V_{\textrm{weak}}(t)= 0.3V_{0}\sin(3\omega t)$, and the red solid curve is for two alternating potentials $V_2(t)= 0.9V_{0}\sin(\omega t)+0.3V_{0}\sin(3\omega t)$.
The frequency is set to $\omega=0.75c^2$. Other parameters are the same as in Fig. \ref{fig3}.}
\end{figure}
In Fig. \ref{fig15} (a), number of electrons as a function of time for different potentials is presented. The frequency is set to $\omega=0.75c^2$. The final number is about $0.07$ for the slow-varying potential, $1.38$ for the fast-varying potential, and $0.77$ for two alternating potentials, respectively. For two alternating potentials, the electron production rate, although high for some time intervals, is followed by a large rate of decline.
Due to the high electric field strength of two alternating potentials, electrons generated under combined potential wells are confined and eventually annihilated with positrons. So the final number of electrons is only half of that under the fast-changing potential.

In Figs. \ref{fig15} (b) and (c), the spacial density and momentum spectra of electrons at the last moment in Fig. \ref{fig14} are presented, respectively. For the low-varying potential, amplitudes of spacial density and momentum spectrum are both smaller than those of the fast-varying potential. This is because the probability of the three-photon process is much smaller than that of the one-photon process. For the fast-varying potential, electrons aggregate in the low momentum region due to the absorption of a supercritical photon by particles in the negative energy continuum, i.e., the direct production of electron-positron pairs by the one-photon process. In addition, since the electric field strength of the fast-varying potential is weak, electrons can cross the potential well to reach outside the potential. So the spatial density is continuously distributed. For two alternating potentials, the momentum spectrum distribution is broadened due to the dynamically assisted effect. The amplitude increases in the high-energy region. The increase in electric field strength causes the low-energy electrons to be bound in the potential well, leading to a decrease in the amplitude in the low-momentum region. So the total number of electrons decreases.

As mentioned above, the combined potentials consisting of a strong low-frequency potential and a weak high-frequency potential are studied, where the high frequency is three times the low frequency.
High-frequency photons from the weak fast-varying potential cooperate with multi-bound states from the strong slow-varying potential to promote the dynamically assisted effect.
In the region of $0<\omega<0.2c^2$, the photon energy provided by the weak high-frequency potential is too low to facilitate the dynamically assisted effect.
In the region of $0.2c^2<\omega<0.66c^2$, there is a significant increase in amplitudes of both the momentum spectrum and the spatial density of electrons duo to the dynamically assisted effect.
When the high-frequency component exceeds the threshold value $2.0c^2$, combined potentials reduce the amplitude of the momentum spectrum in the low-momentum region, leading to a decrease in the total number of electrons.

\subsection{Multiple alternating potentials}
In this section, the enhancement of electron-positron pair generation under multiple alternating potentials is studied.

\begin{figure}[htbp]
\includegraphics[width=10cm]{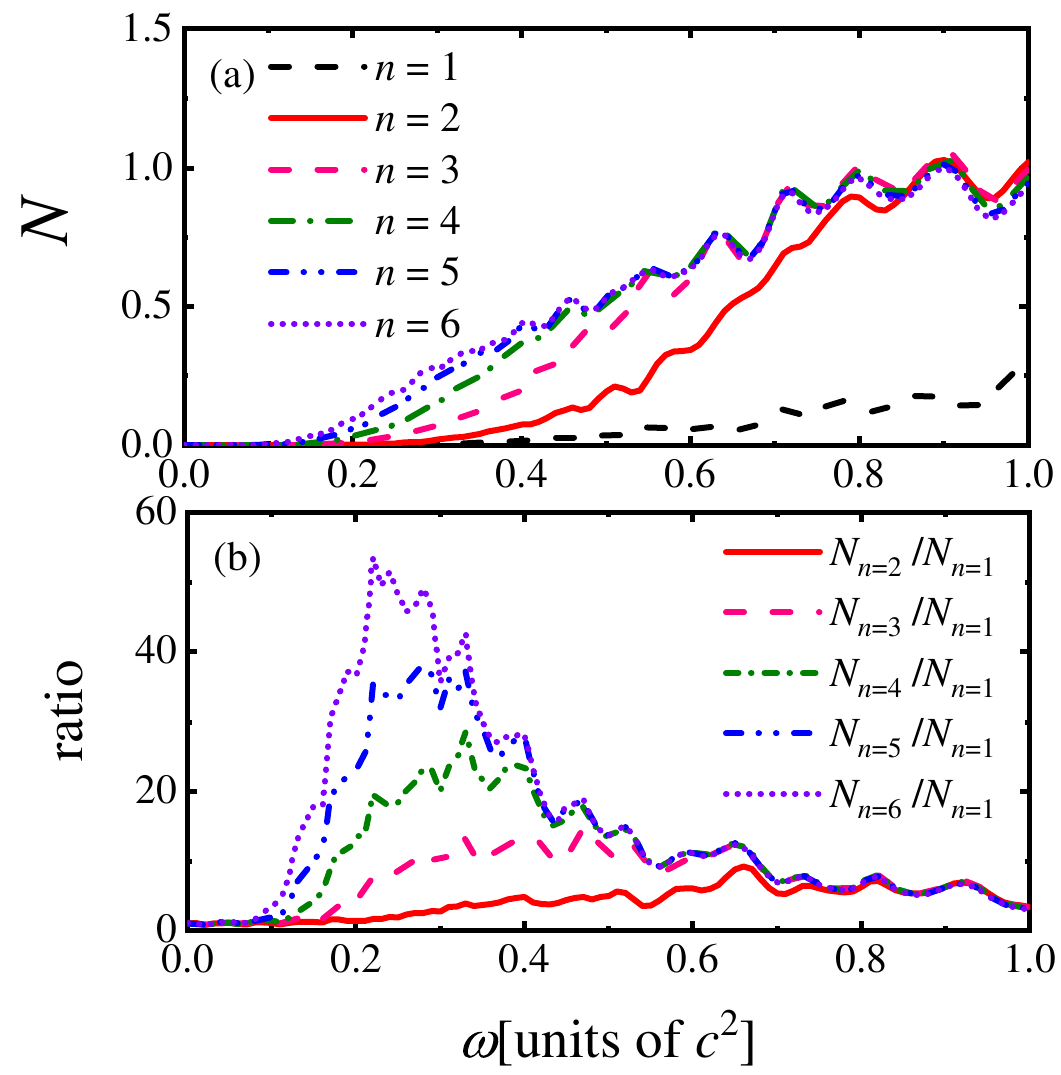}
\caption{\label{fig16} (color online) (a) Number of electrons created under multiple alternating potentials as a function of frequency for $n=1,~2,~3,~4,~5,~6$. (b) Ratio of the number of electrons produced under multiple alternating potentials to that of a single alternating potential. Other parameters are the same as in Fig. \ref{fig3}.}
\end{figure}

In Fig. \ref{fig16} (a), number of electrons created under multiple alternating potentials as a function of frequency for $n=1,~2,~3,~4,~ 5,~6$ is presented, where $n$ is the number of superposed potentials. Other parameters are the same as in Fig. \ref{fig3}. For comparison, when $n=1$ and $2$, the black curve in Fig. \ref{fig3} and the red curve in Fig. \ref{fig8} (a) are shown again in Fig. \ref{fig16} (a).
It is found that number of superposed potentials $n$ has an enhancement effect on the final number of electrons in some frequency regions. For multiple alternating potentials, the frequency component $(2n-1)\omega$ increases gradually with the increase of $n$.
When the frequency $\omega$ is very small, the high-frequency component $(2n-1)\omega$ is also small.
The pair generation process under multiple alternating potentials is still dominated by the Schwinger mechanism. Therefore, the final number of electrons will not increase.
When the high-frequency component $(2n-1)\omega$ is smaller than the threshold value $2.0c^2$, the dynamically assisted effect is enhanced, resulting in an increase in the number of electrons. When the frequency $\omega$ is large enough that the high-frequency component exceeds the threshold value $2.0c^2$, the final number of electrons is reduced. Electrons in the low-momentum region are greatly reduced by combined potentials. This is caused by the phenomenon of high frequency suppression, i.e., the number of electrons created under an oscillation potential decreases when the frequency exceeds $2.4c^2$, which is studied in Ref. \cite{Jiang2013}.

In Fig. \ref{fig16} (b), the ratio of the number of electrons produced under multiple alternating potentials to that of a single alternating potential as a function of frequency is presented. With the increase of frequency, the ratio first increases and then decreases. When $\omega=0.66c^2,~0.47c^2,~0.33c^2,~0.28c^2,~0.22c^2$, the ratio reaches the maximum for $n= 2,~3,~4,~5,~6$, respectively. The maximum frequency $(2n-1)\omega$ are in the range from $2.0c^2$ to $2.5c^2$.
When the maximum frequency $(2n-1)\omega$ exceeds this region, the pair production is suppressed. When $(2n-1)\omega<2.0c^2$, the increase in the number of superposed potentials enhances the generation of electron-positron pairs due to the enhancement of dynamically assisted effect. When the appropriate $n$ and $\omega$ parameters are selected, the number of electrons can be increased by tens of times.

\begin{figure}[htbp]
\includegraphics[width=15cm]{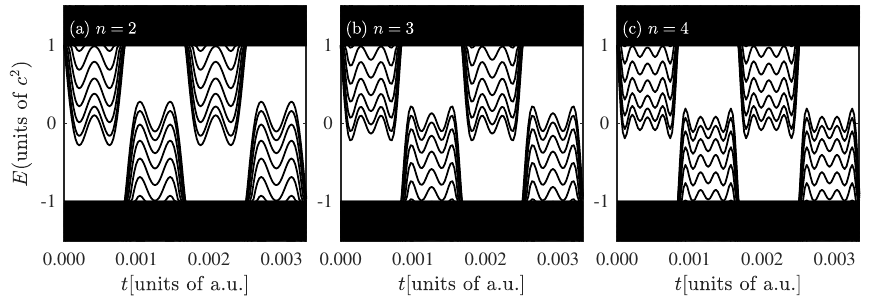}
\caption{\label{fig17} Instantaneous eigenvalues of multiple alternating potentials over time for (a) $n=2$, (b) $n=3$, (c) $n=4$, where $\omega=0.2c^2$. Other parameters are the same as in Fig. \ref{fig3}. }
\end{figure}

Next, we illustrate the enhancement of the pair generation process by superposed potentials at low frequencies.
In Fig. \ref{fig17}, instantaneous energy eigenvalues of multiple alternating potentials varying with time for $n=2,~3,~4$ are presented. The frequency is set to $\omega=0.2c^2$.
Obviously, the number of superposed potentials does not affect the oscillation period, but rather the number of peaks in instantaneous bound states within half an oscillation period.
With the increase of $n$, the effective interaction time of the bound states is increased, which promotes the production of electrons.
In addition, the combination of low-frequency strong fields and high-frequency weak fields greatly increases the production rate.

\begin{figure}[htbp]
\includegraphics[width=15cm]{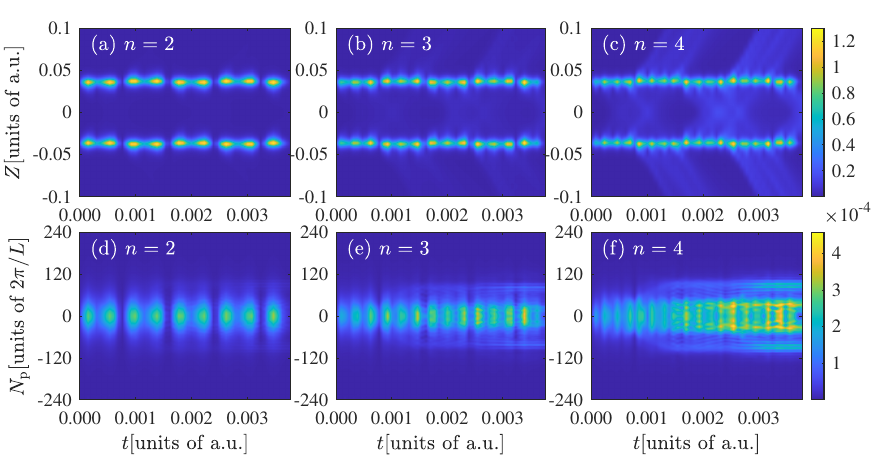}
\caption{\label{fig18} (color online) Time evolutions of the spacial density (top row) and momentum spectrum (bot-
tom row) of electrons created under multiple alternating potentials
for $n=2$ (left column), $n=3$ (middle column), and $n=4$ (right column), where $\omega=0.2c^2$. Other parameters are the same as in Fig. \ref{fig3}.}
\end{figure}
In Fig. \ref{fig18}, time evolutions of the spacial density (top row) and momentum spectrum (bottom row) of electrons created under multiple alternating potentials for $n=2,~3,~4$ are presented.
The frequency is set to $\omega=0.2c^2$. Other parameters are the same as in Fig. \ref{fig3}.
In Figs. \ref{fig18} (a), (b) and (c), electrons are mainly distributed at edges of combined potentials. With the increase of $n$, there are more electrons outside combined potentials.
For $n=2$, two frequency components are set to $\omega=0.2c^2$ and $3\omega=0.6c^2$. The high-frequency component $3\omega$ is relatively small. The Schwinger mechanism dominates the generation of pairs. For $n=3$ and $4$, high-frequency components $5\omega=1.0c^2$ and $7\omega=1.4c^2$ are added. The dynamically assisted effect is enhanced. More electrons are distributed in the high-momentum region.
In Figs. \ref{fig18} (d), (e) and (f), the time effect of momentum spectrum becomes progressively weaker with the increase in the number of superposed potentials.

\begin{figure}[htbp]
\includegraphics[width=15cm]{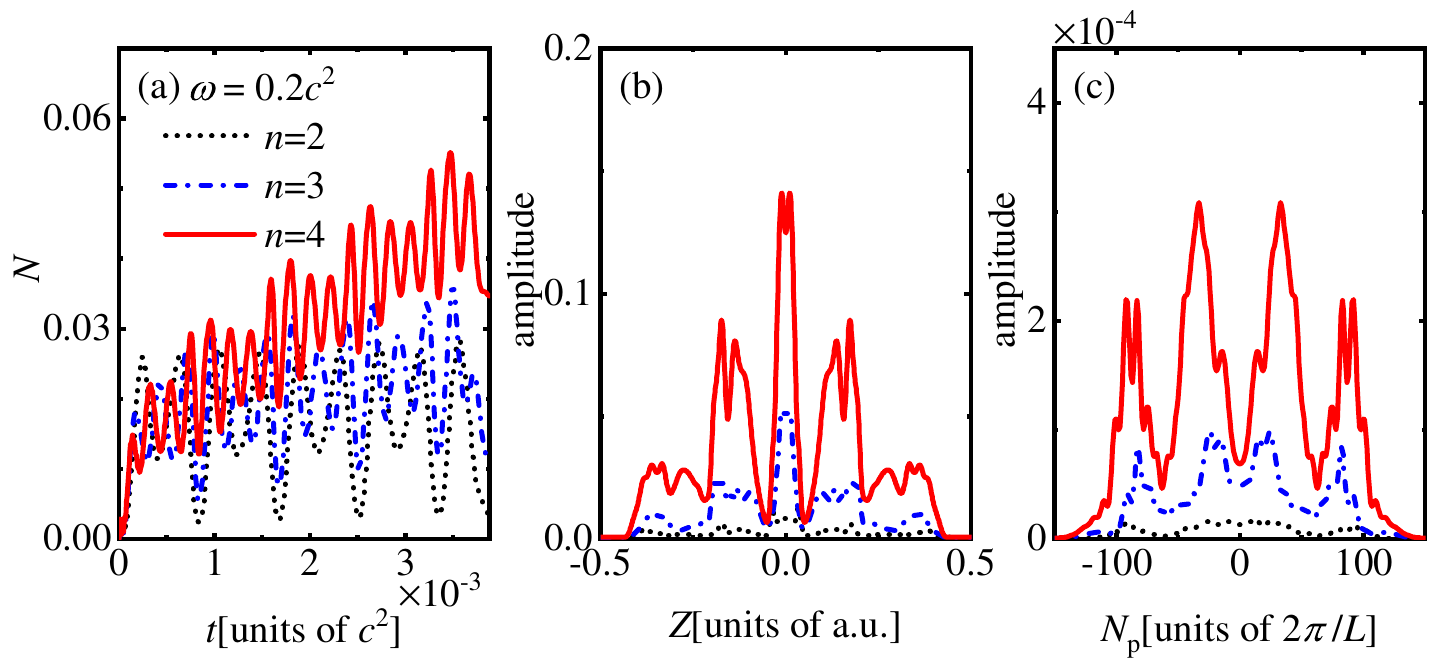}
\caption{\label{Fig19} (color online) (a) Number of electrons as a function of time for $n=2,~3,~4$, where $\omega=0.2c^2$. (b) Spacial density distributions and (c) momentum spectra of electrons at the last moment in Fig. \ref{fig18}. Other parameters are the same as in Fig. \ref{fig3}.}
\end{figure}
In Fig. \ref{Fig19} (a), number of electrons as a function of time for $n=2,~3,~4$ is presented, where $\omega=0.2c^2$.
The number of electrons increases or decreases periodically with time. The number of peaks in each oscillation cycle corresponds to that of superposed potentials. For $n=2$, the final number is very small. With the increase of $n$, the number of electrons oscillates more strongly with time. The final number also gradually increases.

In Figs. \ref{Fig19} (b) and (c), spacial density distributions and momentum spectra of electrons at the last moment in Fig. \ref{fig18} are presented, respectively. With the increase of $n$, amplitudes of spacial density distributions and momentum spectra all increases significantly. In Fig. \ref{Fig19} (b), electrons outside the potential are distributed in steps, especially for $n=3$ and $4$.
It is related to the fact that electrons mainly come out of the potential barriers, as shown in Figs. \ref{fig18} (b) and (c).
In Fig. \ref{Fig19} (c), peak value increases due to the enhanced dynamically assisted effect. A series of peaks are associated with bound states. The high-frequency weak fields provide high-energy photons for the production of electron-positron pairs.
With the increase of $n$, the number of high-energy photons is increased. Therefore, bound states are fully utilized, leading to an increase in peaks of the energy spectrum.

\section{Discussion and Summary}
The enhancement of electron-positron pairs under single, double and multiple alternating potentials is studied by means of computational quantum field theory. The characteristics of electron momentum spectrum and spatial density distribution generated by different mechanisms are studied. The dependence of the electron energy spectrum on instantaneous bound states of the oscillating potential well, potential barrier and the alternating potential is investigated. We focus on comparing the number of electrons produced by two alternating potentials and one alternating potential. The change of generation mechanism caused by the superposition of potential in different frequency regions is analyzed.
Finally, we study the influence of the number of superposed potentials on the pair generation process. The main results are as follows.

1. For an oscillating potential, the momentum spectrum and spacial density of electrons generated under different mechanisms have different characteristics. The electron spatial density distribution and momentum spectrum dominated by Schwinger mechanism have obvious time effect. Electrons are concentrated at the two edges of the potential. In the momentum spectrum, the amplitude is the largest when momentum equals $0$. With the increase of momentum, the amplitude decays exponentially. With the increase of frequency, multiphoton processes dominate the pair generation. More electrons are distributed outside the potential. Multiple peaks appear in the electron momentum spectrum. The momentum spectrum evolves toward a step structure. In addition, the energy level and duration of the bound state affect the position and amplitude of the momentum spectrum peaks, respectively. Electrons produced in the alternating potential are more monoenergetic compared to the oscillating potential well and potential barrier.

2. By combining a deep slow alternating potential with a shallow fast alternating potential, where the frequency of the fast potential is three times that of the slow potential, the number of electrons can be increased several times. For the low frequency case, the number of electrons does not increase under the dominance of the Schiwnger mechanism because the high frequency component is also small. With the increase of frequency, the combination of low-frequency strong fields and high-frequency weak fields enhances the dynamically assisted effect, leading to an increase in the number of electrons.
When the high-frequency component exceeds the threshold value, electrons in the low-momentum region are hindered by the potential well due to the addition of the strong field, leading to a decrease in the final electron number.

3. For multiple alternating potentials, the number of electrons can be increased tens of times by choosing the appropriate frequency.
With the increase of the number of potentials, the high-frequency component gradually increases, which enhances the dynamically assisted effect. When the high-frequency component is greater than the threshold value, continuing to increase the number of potentials has an inhibitory effect on the production of electron-positron pairs.

From the time evolution of the spatial density distribution of electrons, electrons generated in the alternating potential are attracted by the potential well to move inward, which hinders the output of electrons. For a more efficient use of the strong field, the potential barrier model can be considered. We consider combinations of odd multiples of frequencies in order to enhance the effective interaction time of bound states. Although the dynamically assisted effect is enhanced in some frequency regions, different frequency combinations might give better results. In particular, for some chosen parameters, an interesting quasi-monoenergetic structure of the particles produced in the alternating potential does exist compared to that in the oscillating potential well or/and potential barrier, which is believed to be useful to get the high quality positron source.

\begin{acknowledgments}
This work was supported by the Tianyou Youth Talent Lift Program of Lanzhou Jiaotong University Grant No. 1520260417 and the National Natural Science Foundation of China (NSFC) under Grant No. 12375240 and No. 11935008. The computation was carried out at the HSCC of the Beijing Normal University.
\end{acknowledgments}

\begin{appendix}

\section{THE THEORETICAL FRAMEWORK OF COMPUTATIONAL QUANTUM FIELD THEORY}

Let us describe the computational quantum field theory. The time evolution of the operator $\hat{\psi} (z,t)$ for the electron-positron field in a potential $V(z,t)$ is given by the Dirac equation\cite{Krekora2005},
\begin{equation}\label{Eq Dirac}
i\partial \hat{\psi} \left(z,t\right) / \partial{t}=\left[c\alpha_z \hat{P}_z+\beta c^2+V\left(z,t\right)\right] \hat{\psi}\left(z,t\right),
\end{equation}
where $\alpha_z$ and $\beta$ are Dirac matrices, $c$ is the speed of light in vacuum, and $V\left(z,t\right)$ is external field that varies with time $t$ in the $z$ direction. We use atomic units (a.u.) as $\hbar=e=m_e=1$.
By introducing the creation and annihilation operators, the field operator $\hat{\psi}(z,t)$ can be decomposed as follows:
\begin{equation}\label{Eq Field Operator}
\begin{aligned}
\hat{\psi}(z,t)&=\sum_{p}\hat{b}_p(t)W_p(z)+\sum_{n}\hat{d}_n^{\dag}(t)W_n(z) \\
&=\sum_{p}\hat{b}_pW_p(z,t)+\sum_n\hat{d}_n^\dag W_n(z,t),
\end{aligned}
\end{equation}
where $p$ and $n$ denote the momenta of positive- and negative-energy states, $\sum_{p(n)}$ presents the summation over all states with positive $($negative$)$ energy, $W_{p}(z)=\langle z|p\rangle$ ($W_{n}(z)=\langle z|n\rangle$) is the field-free positive (negative) energy eigenstate. Note that $W_{p}(z,t)=\langle z|p(t)\rangle$ and $W_{n}(z,t)=\langle z|n(t)\rangle$ satisfies the single-particle time-dependent Dirac equation (\ref{Eq Dirac}).
From Eq.(\ref{Eq Field Operator}), we obtain
\begin{equation}\label{Eq fermion operators}
\begin{aligned}
\hat{b}_p(t)&=\sum_{p'}\hat{b}_{p'}U_{pp'}(t)+\sum_{n'}\hat{d}_{n'}^\dag U_{pn'}(t),\\
\hat{d}_n^\dag(t)&=\sum_{p'}\hat{b}_{p'}U_{np'}(t)+\sum_{n'}\hat{d}_{n'}^\dag U_{nn'}(t),\\
\hat{b}_p^\dag(t)&=\sum_{p'}\hat{b}_{p'}^\dag U_{pp'}^*(t)+\sum_{n'}\hat{d}_{n'}U_{pn'}^*(t),\\
\hat{d}_n(t)&=\sum_{p'}\hat{b}_{p'}^\dag U_{np'}^*(t)+\sum_{n'}\hat{d}_{n'}U_{nn'}^*(t),
\end{aligned}
\end{equation}
where $U_{pp'(t)}=\langle p|\hat{U}(t)|p'\rangle$, $U_{pn'(t)}=\langle p|\hat{U}(t)|n'\rangle$, $U_{nn'(t)}=\langle n|\hat{U}(t)|n'\rangle$, and $U_{np'(t)}=\langle n|\hat{U}(t)|p'\rangle$ and the time evolution propagator $\hat{U}(t)=\textrm{exp}\{{-i\int^t d\tau [c\alpha_z \hat{p}_z+\beta c^2+V(z,\tau)]}\}$.

In Eq. (\ref{Eq Field Operator}), the electronic portion of the field operator is defined as $\hat{\psi}_e(z,t)\equiv \sum_p \hat{b}_p(t)W_p(z)$. So we can obtain the probability density of created electrons by
\begin{equation}\label{Eq Density}
\begin{aligned}
\rho(z,t)&=\langle \text{vac}|\hat{\psi}_e^\dag (z,t)\hat{\psi}_e(z,t)|\text{vac}\rangle \\
&=\sum_n \left |\sum_p U_{pn}(t)W_p(z)\right |^2.
\end{aligned}
\end{equation}
By integrating this expression over space,~the number of created electrons can be obtained as
\begin{equation}\label{Eq Number}
N(t)=\int \rho(z,t)dz=\sum_p \sum_n \left|U_{pn}(t)\right|^2.
\end{equation}
The time evolution propagator $U_{pn}(t)$ can be numerically calculated by employing the split-operator technique\cite{Tang2013}. Therefore, according to Eqs. (\ref{Eq Density}) and (\ref{Eq Number}), we can compute various properties of the electrons produced under the action of the external potential.

\end{appendix}

\end{document}